\documentclass[aps,pra,twocolumn,showpacs,groupedaddress,nofootinbib]{revtex4-1}

\usepackage[utf8]{inputenc}  %Input what you want e.g., é, ł, a, ü
\usepackage[T1]{fontenc}     %Output what you want e.g., é, ł, a, ü
\usepackage[colorlinks,linkcolor=magenta,urlcolor=blue]{hyperref}  %Hyperlinks (pink, green, blue)
\usepackage{graphicx} % Package to insert exteral figures
\usepackage[babel]{microtype}  %Improves text justification
\usepackage{amsmath,amssymb,amsthm,bm,mathtools,amsfonts,mathrsfs,bbm} %Usefull math packages
\usepackage{xspace}  %Useful to add space in macros
\usepackage{centernot}
\usepackage{sistyle}
\usepackage{gensymb}
\usepackage{natbib}
\usepackage{dsfont}
\usepackage{epsfig}
\usepackage{tabularx}
\usepackage{hyperref}

\usepackage[table]{xcolor}
\usepackage{colortbl}
\usepackage{color}
\usepackage{multirow}
\usepackage{hhline}
\usepackage{tabu}
\usepackage{epstopdf}
\usepackage{booktabs}
\usepackage{textgreek}
\newcommand{\bra}[1]{\left\langle #1 \right|}
\newcommand{\ket}[1]{\left| #1 \right\rangle}

\newcommand{\exval}[3]{\left\langle #1 \middle| #2 \middle| #3 \right\rangle}
\newcommand{\abs}[1]{\left|#1\right|}

\def\be{\begin{eqnarray}}
\def\ee{\end{eqnarray}}
\mathtoolsset{centercolon}

\usepackage{changes}
\usepackage{comment}

\newcommand{\dt}[3]{\cellcolor{#1} \scriptsize #2(#3)}

\newcommand{\dtt}[1]{\scriptsize #1}
\definecolor{gray4}{gray}{0.8}
\definecolor{gray2}{gray}{0.6}

\begin{document}

\title{Quantifying photonic high-dimensional entanglement}

\author{Anthony Martin$^1$}
\author{Thiago Guerreiro$^{1}$}
\altaffiliation{Present address: Departamento de F\'isica, Pontif\'icia Universidade Cat\'olica, PUC-Rio, Pr\'edio Leme, Sala 652 L, G\'avea, 22451900 Rio de Janeiro, RJ Brasil}
\author{Alexey Tiranov$^1$}
\author{S\'ebastien~Designolle$^1$}
\author{Florian Fr\"owis$^1$}
\author{Nicolas Brunner$^1$}
\author{Marcus Huber$^{1,2}$}
\author{Nicolas Gisin$^1$}
%\author{Hugo Zbinden$^1$}
\affiliation{$^1$Group of Applied Physics, University of Geneva, CH-1211 Geneva 4, Switzerland}
\affiliation{$^2$Institute for Quantum Optics and Quantum Information, Austrian Academy of Sciences, A-1090 Vienna, Austria}
\date{\today}

\begin{abstract}
High-dimensional entanglement offers promising perspectives in quantum information science. In practice, however, the main challenge is to devise efficient methods to characterize high-dimensional entanglement, based on the available experimental data which is usually rather limited. Here we report the characterization and certification of high-dimensional entanglement in photon pairs, encoded in temporal modes. Building upon recently developed theoretical methods, we certify an entanglement of formation of 2.09(7) ebits in a time-bin implementation, and 4.1(1) ebits in an energy-time implementation. These results are based on very limited sets of local measurements, which illustrates the practical relevance of these methods. 
\end{abstract}

\maketitle

%\section{Introduction}

Entanglement is among the most fascinating features of quantum theory and at the heart of quantum information processing. 
In recent years, a growing interest has been devoted to the possibility of generating entangled states of high-dimensions. Such states can in principle contain a large amount of entanglement, which is conceptually interesting but also offers novel perspectives for applications in quantum information,
%~\cite{Raussendorf2001}
particularly in quantum communications~\cite{Bennett1999,Bechmann-Pasquinucci2000,Cerf2002,Sheridan2010}.

Several experimental platforms have been considered for the creation of highly entangled states, in particular in photonics. These include encodings based on energy-time~\cite{Franson1989,Richart2012,Thew2004}, time-bins~\cite{Brendel1999,DeRiedmatten2002,Ikuta2016,Stucki2005}, orbital angular momentum~\cite{Mair2001,Dada2011,Krenn2014}, and frequency modes~\cite{Olislager2012a,Bernhard2013,Jin2016}. Thus highly entangled states are now routinely created in all of these platforms. Also the entanglement of these states can be detected experimentally, via the use of entanglement witnesses or Bell inequalities \cite{Pan2012,Brunner2014}. 

However, the real challenge in this area is the experimental certification of large entanglement, such as high-dimensional entanglement. That is, not only to certify the mere presence of entanglement, but to provide an actual certification of the amount of entanglement present in the state. This issue is challenging for two different reasons. 

First, the characterization of a quantum state of high dimension via standard methods (e.g. quantum tomography) typically requires the estimation of a considerable number of independent parameters, which in turn requires a large number of different measurements to be performed. In practice this is extremely cumbersome and essentially infeasible. Alternative approaches have been developed \cite{Gross2010}, requiring much less measurements to be performed, but are usually based on some extra assumptions on the state (for instance, the state being of high purity). 

Second, the measurements that can actually be performed in a real experiment are typically limited. Therefore, the practical use of efficient methods to certify high-dimensional entanglement is further constrained by a restriction on the class of measurements available in the lab. This has been an active area of research in recent years \cite{Giovannini2013,Tonolini2014,Howland2016,Erker2016,Tiranov2016}, motivated by the prospects of applications in quantum communications \cite{Mirhosseini2015,Zhong2015}.

Here we report the characterization and certification of high-dimensional entanglement in photonic systems, based on very sparse experimental data. To do so, we build upon the theoretical methods recently developed in Ref. \cite{Alexey2016}. We first discuss a time-bin entangled two-photon experiment, in which we certify presence of at least 2.09~(7) ebits of entanglement of formation. In other words, we certify that the created states contains (i) an amount of entanglement equivalent to more than two maximally entangled two-qubit pairs, and (ii) entanglement in (at least) $5 \times 5$ dimensions. Second, we report an experiment using energy-time entangled photon pairs, and certify up to 4.1~(1) ebits of entanglement of formation, based on few additional assumptions. This certifies entanglement in a state of dimension of (at least) $18 \times 18$. To the best of our knowledge, this represents the highest values of entanglement of formation certified so far in any experiment. These results demonstrate the potential of temporal entanglement as a platform for creating and certifying quantum states featuring a large amount of entanglement. 

\begin{figure}[t!]
\includegraphics[width=1\columnwidth]{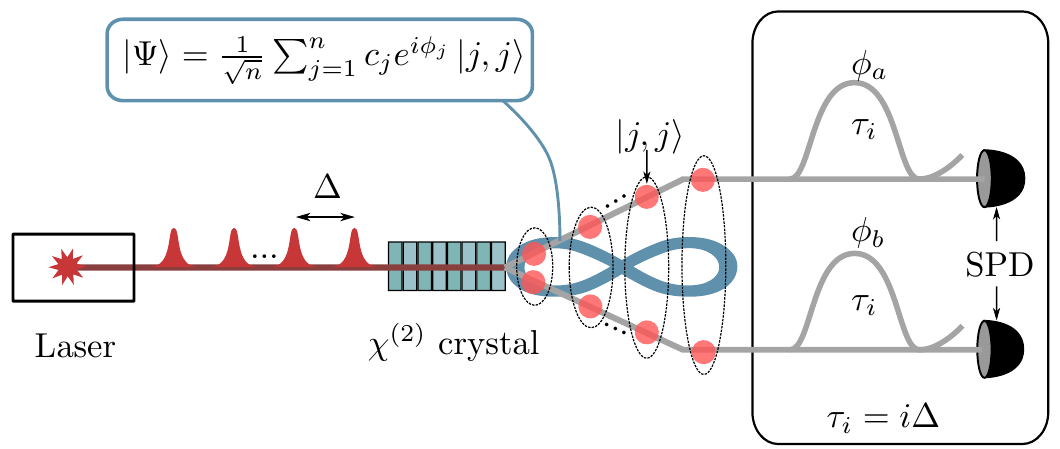}
\caption{Schematic view of the experiment for creating and quantifying high-dimensional entanglement using limited measurement data. The time-bin entangled two-photon state is produced by spontaneous parametric down conversion process (SPDC) using a $\chi^{(2)}$ nonlinear crystal. The crystal is pumped by a mode-locked laser with high repetition rate producing a photon pair in well defined temporal modes $\ket{j,j}$. A Franson type setup \cite{Franson1989} consisting of two interferometers is used to analyse the resulting state and reveal its entanglement. 
%b) Coincidences measured by the single photon detectors (SPD) are used to extract visibilities $V_i$ for given interferometer delay $\tau_i$. Extracted values represent different diagonals in reconstructed density matrix $\rho$.
\label{fig_principle}}
\end{figure}

\emph{Setup.---} The setup of our experiment is sketched in \figurename{~\ref{fig_principle}}. Time-bin entanglement is generated using spontaneous parametric down-conversion (SPDC). A picosecond mode locked laser at 532 nm creates a train of pulses, which then stimulates a type 0 periodically poled lithium niobate crystal to generate photon pairs at a wavelength of 810~nm and 1550~nm. The delay between two successive pulses is $\Delta = 2.3$~ns. At the output of the crystal, when we have exactly one pair, the state generated is of the form: 
\begin{equation}\label{eq_state}
\ket {\Psi} = \frac{1}{\sqrt{n}} \sum_{j=1}^{n} c_j e^{ i \phi_j} \ket{j,j}
\end{equation}
where $\ket{j,j}$ denotes the states where both photons are in the pulse $j$, which has amplitude $c_j$, and phase $\phi_j$. The mode-locked laser preserves the amplitude and phase relation over a large number of pulses, $n$. This means that $c_j$ and $\phi_j$ can be consider constant for $n \Delta$ smaller than the coherence time of the laser. In our configuration based on a laser with a coherence time greater than 1 \textmu s and a repetition rate of 430~MHz, the coherence of this state can be preserved for $n \leq 400$. Moreover, the pump power is set in such a way that the probability to generate two photon pairs in a n-pulse train is negligible.  

The created time-bin entangled two-photon state is then analyzed. First, the two photons (of each pair) are separated by a dichroic mirror and each photon is sent to a bulk unbalanced interferometer. The delay between the short and long arms of the interferometers can be set to $\Delta $ and $2\Delta$ in order to analyse the coherence between two neighbor ($j$ and $j+1$) and next-neighbor ($j$ and $j+2$) temporal modes. These delays are much larger than the pulse duration of the laser $\tau_p \approx 10$~ps and the coherence time of the down-converted photons $\tau_c \approx 1$~ps (the coherence time of the photons is estimated from the bandwidth of the photon at 1550~nm which is around 3~nm). In this case, only second order interference can be measured by analysing the coincidence rate at the output of the interferometers, which correspond to the local projections onto the state $\bra{j+i,j+i}e^{-i (\phi_a + \phi_b)} + \bra{j,j}$ (with $i=1,2$), where $\phi_a$ and $\phi_b$ correspond to the relative phase between the two arms of the interferometers. These phases can be adjusted by piezo actuators. To extract the visibility, the phase of one of the two interferometers is scanned to find the maximum and minimum coincidence rates, which correspond to constructive and destructive interference, respectively. At the output of each interferometer, the photons are detected via a single photon detector (SPD), based on an silicon (resp. InGaAs) avalanche photodiode for the photon at 810~nm (resp. 1550~nm). To associate the detections with the correct temporal modes, the detection events are sent to a time-to-digital converter where the clock is set on the laser frequency divided by $2^{12}$. More precisely, the temporal mode $j$ corresponding to each detection is defined from the time delay between the clock trigger and the detection event. 

\emph{Entanglement certification.---} Our goal here is to characterize the entanglement of the time-bin entangled state we create. This is however a nontrivial problem, due to the very limited data available from the experiment. In particular, we cannot reconstruct the full density matrix $\rho$ of the state, due to the fact that we are not able to experimentally measure each element $\exval{j,k}{\rho}{j',k'}$. This would require having basically $n$ unbalanced interferometers, which is clearly unpractical.

Specifically, our setup allows us to measure only the following quantities. First, we can measure coincidence events in the time-of-arrival basis, which gives access to the diagonal density matrix elements $\exval{j,k}{\rho}{j,k}$. Second, we can measure the interference visibility between two neighboring temporal modes ($j$ and $j+1$), and similarly for two next-neighboring temporal modes ($j$ and $j+2$). Hence we can estimate the off-diagonal elements $\exval{j,j}{\rho}{j+i,j+i}$ for $i=1,2$ (\figurename{~\ref{fig_principle}}). Apart from these quantities, we cannot get access to any further elements of~$\rho$.

Although this data is rather limited, it turns out that we can nevertheless efficiently characterize the entanglement produced by the source. In particular we obtain strong lower bounds on the amount of entanglement contained in the state. To do so we build upon recent theoretical methods presented in Ref. \cite{Alexey2016}. 
\begin{figure*}
\begin{tabular}{ll}
a) & b)\\
\includegraphics[width=\columnwidth]{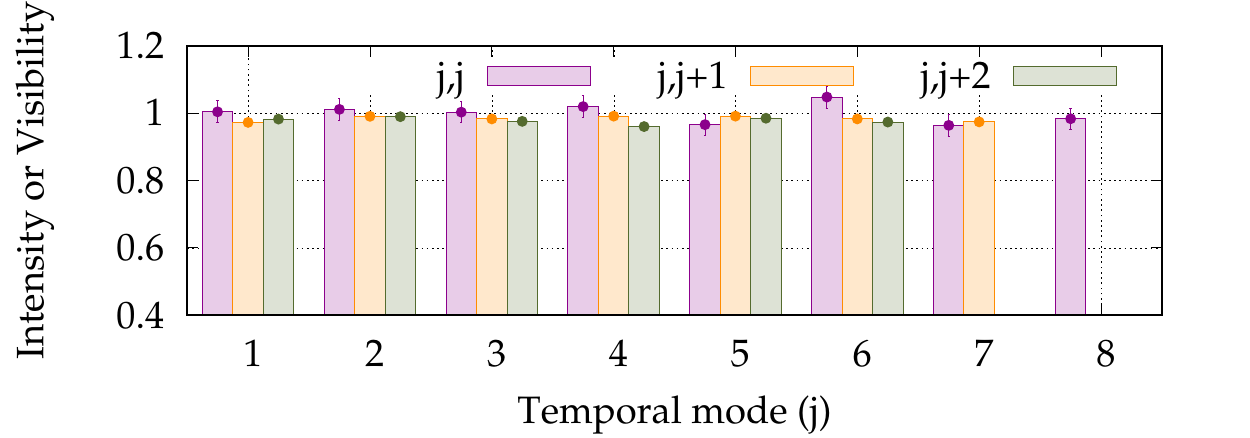}&
\includegraphics[width=\columnwidth]{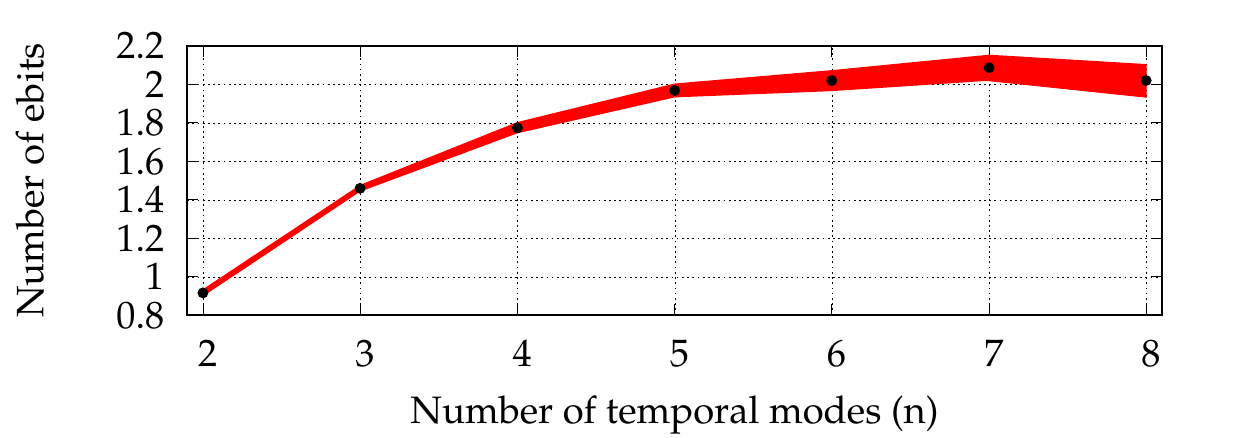}
\end{tabular}
\caption{\label{pulsed_vis} Results for the time-bin experiment. 
a) Here $(j,j)$ corresponds to the intensity probabilities ($|c_j|^2$ in Eq. (\ref{eq_state}) ), while $(j,j+1)$ and $(j,j+2)$ represent the visibilities of the two-photon interference between two neighboring and next-neighboring temporal modes. We consider 8 temporal modes. 
b) Lower bounds on the entanglement of formation (in terms of ebits) as a function of the number of temporal modes $n$ considered. The data allows us to certify an entanglement of formation of  2.09(7) ebits, taking $n=7$. The black dots correspond to the mean number of ebits and the one $\sigma$ statistical error is represented by the filled red curve.
}
\end{figure*} 
More specifically, this approach will allow us to lower-bound the entanglement of formation of $\rho$, $E_{oF} $. This measure represents the minimal number of "ebits" (i.e. the number of maximally entangled two-qubit states) required in order to produce $\rho$ via an arbitrary LOCC procedure \cite{Horodecki_review}. This measure thus has a clear operational meaning. Specifically, we have that 
\begin{equation} \label{EoF}
E_{oF} \geqslant -\log_2(1-\frac{B^2}{2})
\end{equation}
where the quantity $B$ is defined as:
\begin{multline}
B =  \frac{2}{\sqrt{|C|}} \left( \sum_{\underset{j<k}{(j,k) \in C}} \abs{\exval{j,j}{\rho}{k,k}} \right.\\
\left. - \sqrt{ \exval{j,k}{\rho}{j,k} \exval{k,j}{\rho}{k,j}}  \right).
\label{eq_B}
\end{multline}
Here $|C|$ denotes the cardinality of the set $C$, i.e. the number of pairs of indices $(j,k)$ to be considered in the sum. By taking more and more elements in $C$, one obtains typically better bounds on $E_{oF}$. How many pairs of indices can be considered depends on how many off-diagonal elements of $\rho$ are known. Note that $B$ provides a lower-bound on the concurrence of $\rho$~\cite{Huber2013}.

While the data available in our experiment does not allow us to reconstruct the complete density matrix, we can nevertheless get strong bounds on the entanglement of formation of $\rho$. First, note that the terms $\exval{j,k}{\rho}{j,k}$ (with $j<k$) in Eq. \eqref{eq_B} are directly related to the coincidence to accidental ratio (CAR) of the source, which can be measured. Concerning the off-diagonal terms $\exval{j,j}{\rho}{k,k}$, recall that we can only determine directly those for which $| k-j | \leq 2$. In Ref. \cite{Alexey2016}, it was shown that the remaining unknown off-diagonal elements can be efficiently lower-bounded by focusing on certain submatrices of $\rho$, and imposing their positivity (which follows from the positivity of $\rho$). Here we combine these ideas with semi-definite programming (SDP) techniques. Specifically, we focus on the sub-matrix of $\rho$: $\tilde{\rho}_{jk} = \Re(\exval{j,j}{\rho}{k,k}) $ \footnote{Note that $\tilde{\rho}$ is usually sub-normalized in practice, due to noise terms. It will then be convenient to renormalize $\tilde{\rho}$.}. Via SDP\footnote{We used the yalmip interface and the sedumi solver \cite{yalmip,sedumi}, which allow one to define an objective function as the sum of absolute values of matrix elements.}, one can then minimize the expression $\sum_{j<k} \abs{\exval{j,j}{\tilde\rho}{k,k}}$ (for $(j,k)$ in the set $C$), under the constraints that $\tilde{\rho}$ is positive and that some elements of $\tilde{\rho}$ (or some linear combinations of them) are known from the data.
From the result of this SDP, we then get a lower bound on the entanglement of formation, via Eqs. \eqref{eq_B} and \eqref{EoF}. Importantly, the bound we obtain here is essentially tight, given that the solution returned by the SDP is a valid density matrix. Among all possible physical states compatible with our data, the SDP solution corresponds to the one featuring the smallest amount of entanglement. 

\begin{table}[b!]
\begin{tabular}{c|@{\extracolsep{\fill}} *{10}{c@{\extracolsep{\fill}}}}
			& \dtt $\ket{1,1}$ 	&  \dtt $\ket{2,2}$ 	& \dtt $\ket{3,3}$	& \dtt $\ket{4,4}$	& \dtt $\ket{5,5}$	&\dtt $\ket{6,6}$ 	&\dtt $\ket{7,7}$ 	& \dtt $\ket{8,8}$  \\ \hline
 \dtt $\bra{1,1}$ &\dt{gray4}{1.01}{4} & \dt{gray4}{0.98}{2} &  \dt{gray4}{0.99}{2} & \dt{white}{0.96}{3} & \dt{white}{0.91}{3} & \dt{white}{0.89}{4} & \dt{white}{0.84}{4} & \dt{white}{0.75}{5} \\
 \dtt $\bra{2,2}$ &\dt{gray4}{0.98}{3} & \dt{gray4}{1.02}{3} &  \dt{gray4}{1.00}{2} & \dt{gray4}{1.01}{2} & \dt{white}{0.96}{3} & \dt{white}{0.97}{3} & \dt{white}{0.91}{3} & \dt{white}{0.84}{4} \\
 \dtt $\bra{3,3}$ &\dt{gray4}{0.99}{2} & \dt{gray4}{1.00}{2} &  \dt{gray4}{1.00}{3} & \dt{gray4}{1.00}{2} & \dt{gray4}{0.96}{2} & \dt{white}{0.96}{3} & \dt{white}{0.91}{3} & \dt{white}{0.85}{4} \\
 \dtt $\bra{4,4}$ &\dt{white}{0.96}{3} & \dt{gray4}{1.01}{2} &  \dt{gray4}{1.00}{2} &  \dt{gray4}{1.02}{3} &  \dt{gray4}{0.98}{2} &  \dt{gray4}{1.00}{2} &  \dt{white}{0.95}{3} &  \dt{white}{0.90}{3} \\
 \dtt $\bra{5,5}$ &\dt{white}{0.91}{3} & \dt{white}{0.96}{3} &  \dt{gray4}{0.96}{2} &  \dt{gray4}{0.98}{2} &  \dt{gray4}{0.96}{3} &  \dt{gray4}{1.00}{2} &  \dt{gray4}{0.95}{2} &  \dt{white}{0.92}{3} \\
 \dtt $\bra{6,6}$ &\dt{white}{0.89}{4} & \dt{white}{0.96}{3} &  \dt{white}{0.96}{3} &  \dt{gray4}{1.00}{2} &  \dt{gray4}{1.00}{2} &  \dt{gray4}{1.05}{3} &  \dt{gray4}{0.99}{2} &  \dt{gray4}{0.99}{2} \\
 \dtt $\bra{7,7}$ &\dt{white}{0.84}{4} & \dt{white}{0.91}{3} &  \dt{white}{0.91}{3} &  \dt{white}{0.95}{3} &  \dt{gray4}{0.95}{2} &  \dt{gray4}{0.99}{2} &  \dt{gray4}{0.96}{3} &  \dt{gray4}{0.95}{2} \\
 \dtt $\bra{8,8}$ &\dt{white}{0.75}{5} & \dt{white}{0.84}{4} & \dt{white}{0.85}{4} & \dt{white}{0.90}{3} & \dt{white}{0.99}{3} &  \dt{gray4}{0.99}{2} &  \dt{gray4}{0.95}{2} &  \dt{gray4}{0.98}{3} \\
\end{tabular}
\caption{\label{tab_rho} Sub-matrix $\tilde \rho$ outputted by the SDP procedure. The gray cells are the elements experimentally measured while the white cells are a priori unknown. The values are normalized with respect to the dimension of $\tilde \rho$. Error bars are estimated over 20 experimental runs for gray cells, and via Monte Carlo simulation for white cells.}
\end{table}

\begin{figure*}[t!]
\begin{tabular}{l l }
a) & b) \\
\includegraphics[width=\columnwidth]{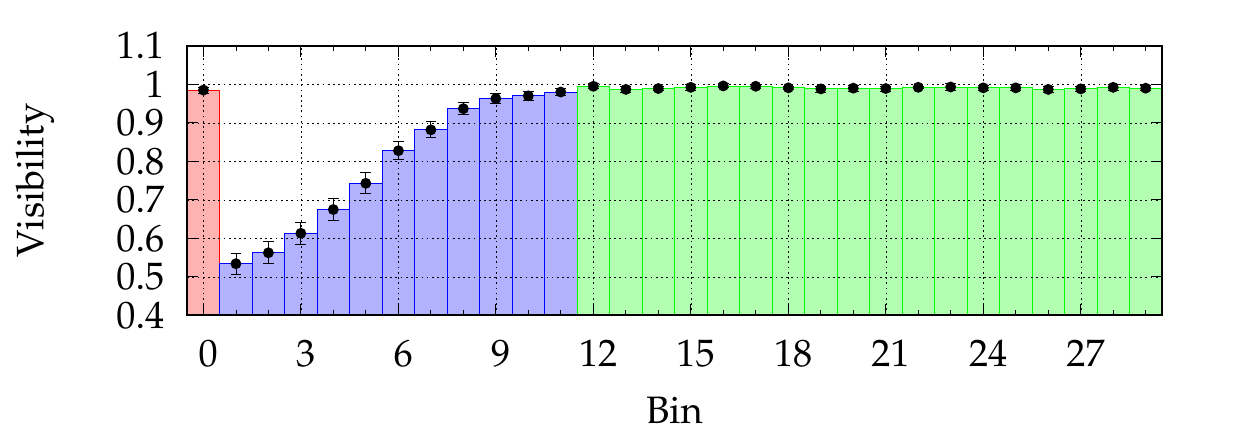} &
\includegraphics[width=\columnwidth]{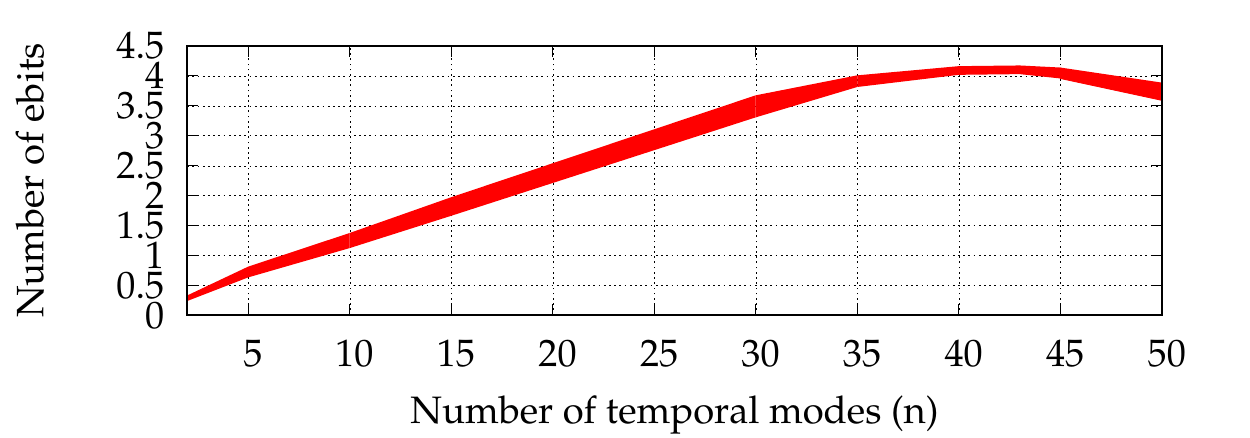}
\end{tabular}
\caption{
Results of the energy-time experiment. a) Visibility of the two-photon interference representing coherence between temporal modes $\ket{j,j}$ and $\ket{j+n,j+n}$ where $n$ is a bin number. The bin $n=0$ corresponds to the first order interference with visibility of 98(1)\%. Up to bin number 11 (blue), the visibility is rather low due to the jitter of the detection system. For the remaining bins (green), the average visibility is 98.5(8)\%. In order to lower bound the entanglement of formation, we use bins 12 to 29.
b) Lower bounds on the entanglement of formation (in terms of ebits) as a function of the number of temporal modes $n$ for the CW experiment. Energy-time entanglement containing up to 4.1(1) ebits is certified for $n=42$. The filled red curve represents the one $\sigma$ statistical error.
\label{fig_CW_vis}}
\end{figure*}

\emph{Experimental results.---} Next we apply the above methods to our experimental data. We consider 8 temporal modes. Measurements in the time-of-arrival basis lead to coincidences events, see \figurename{~\ref{pulsed_vis}.(a)}. This also allows to estimate the CAR to be $>10^3$. We thus take $\exval{j,k}{\rho}{j,k}  = 10^{-3}$ for any $k \neq j$. Also, using the unbalanced interferometers we measure the interference visibilities by applying the local projections $\bra{j+i,j+i}e^{i (\phi_a + \phi_b)} + \bra{j,j}$ (for $i=1,2$ and $j=1,...,8-i$) and vary the relative phase. As seen from \figurename{~\ref{pulsed_vis}.(a)}, the average visibility is $ \sim 98\% $.

Next we run the SDP procedure explained above. In order to find the largest value of the entanglement of formation, it is useful to consider all contiguous subsets of $n=2,...,8$ of the total 8 temporal modes. Increasing $n$ may increase the first term of the quantity $B$ (see Eq. \eqref{eq_B}), but will also decrease the norm of each term due to the normalization of the state. In practice we scan over all possible values of $n$ and keep the best. Moreover, one should tune the subset $C$. Here, enumerating all $2^{n(n-1)/2}$ is not possible; in practice we simply take $C$ to contain all possible pair of indices $(j,k)$, thus taking all off-diagonal elements into account. The largest value is found for $n=7$: $E_{oF} = 2.09 (7) $ ebits; see \figurename{~\ref{pulsed_vis}.(b)}. This result also certifies that our time-bin entangled states is of dimension at least $5 \times 5$. Table 1 shows the sub-matrix $\tilde{\rho}$ outputted by the SDP, highlighting known and unknown matrix elements.

\emph{Energy-time entanglement.---} Our theoretical methods can in principle be used for certifying much higher entanglement. Experimentally, this amounts to consider more temporal modes, and measure more off-diagonals elements. However, in our configuration, this would require to increase the path difference of the interferometers up to 70~cm, which is unpractical. 

Instead, we pursue a different approach. In order to effectively reduce the interferometer size one can modify the source of entanglement, using e.g. a pulsed laser with a faster repetition rate or a continuous laser. Here we explore the second option. In this case, the entanglement originates from energy-time correlations induced by the narrow spectrum (long coherence time) of the pump laser. The resulting state can nevertheless be equivalently written in the form of Eq.\eqref{eq_state} by properly defining the temporal modes.

In practice we used another source of photon pairs based on a type-0 pigtail PPLN waveguide stimulated by a diode laser at 780~nm to generate degenerate photons at 1560~nm. To filter down the photons to 100~GHz, the output of the source is connected to a 200~GHz dense wavelength division multiplexer, which increases the coherence time of the photons to 4.4~ps. To analyze the coherence between different temporal modes the photon pairs are sent to a single folded bulk Franson interferometer~\cite{Thew2002} and a single photon detector is placed at each output port to measure the second order correlation. A motorised mirror placed on one arm can continuously change the interferometer delay length from 0 to 29~cm.

The main difference between this approach and our first experiment is coming from the fact that we can now choose the delay between two adjacent temporal modes (given previously by $\Delta$) as long as the modes do not overlap, i.e. that our entangled state satisfies $ \exval{j,k}{\rho}{j,k} \approx 0$ $\forall j \neq k$. To do so, we fix the time between two adjacent modes equal to 33~ps (1~cm), which allows us to analyse 29 temporal modes. In this configuration, the temporal jitter of the detection scheme (around 200\,ps) is larger than the delay between two temporal modes, which means that we cannot access each mode individually anymore.

As in our first experiment, the values of the off-diagonal elements are directly related to the visibility of the two-photon interference. As shown in \figurename{~\ref{fig_CW_vis}.(a)}, when the delay between the two arms is set to zero, $i = 0$, we observe single photon interferences with a visibility of 98.5(8)\%. When we increase the delay length from 1 to 11 cm (33\,ps to 337\,ps), we observe that the visibility drops to 53(3)\% and increases to 98(1)\%. For these delays, we do not observe single photon interferences anymore, and two-photon interference is limited by the detection system. Indeed, due to the jitter of the detection scheme, the cases where the two photons pass through different arms cannot be removed by a temporal post-selection. For $i>11$, we observe that the visibilities are constant, and around 99\%. In this configuration, we cannot infer the values of elements of the 11 first off-diagonals, but only from the off-diagonals 12 to 29. Also, we measure a CAR of 5650 using low noise detectors cooled down to -90$^\circ$C with a Stirling cooler~\cite{Korzh2014}.

In order to quantify entanglement from this data, we use again our theoretical method and analyze the sub-matrix $\tilde\rho$. Contrary to our first experiment, the elements of $\tilde\rho$ are not known individually, but only certain averages. In particular, the measured visibilities provide the following constraints: $ \frac{1}{n-i}\sum_{j=1}^{n-i} \Re(\exval{j,j}{\tilde\rho}{j+i,j+i}) = V_i$ for $i\in \{12,29\}$, which are included in the SDP. We also have a constraint on the diagonal elements of $\tilde\rho$, namely $\frac{1}{n}\sum_{j=1}^{n}\exval{j,j}{\tilde\rho}{j,j} = 1$ which follows from normalization of the state $\rho$.
%we can assume that all modes are equally distributed, i.e. $\exval{j,j}{\tilde\rho}{j,j} = 1$. This is well justified by the fact that the laser has a line width of 1~MHz, such that the power is essentially constant over 30,000 modes. 
All remaining matrix elements are unknown, hence the sub matrix $\tilde\rho$ takes the form:
%In the SDP, we now use the (non-normalized) sub-matrix $\tilde\rho$, the elements of which are: $\tilde\rho_{j,j} = 1$ and $\tilde\rho_{j,j+i} = \tilde\rho_{j+i,j} = V(i)$ for all $j$ and $i\in[12,29]$, while all remaining elements are unknown, which gives a sub-matrix of the form: \\
\begin{center}
\includegraphics [width = 0.7\columnwidth]{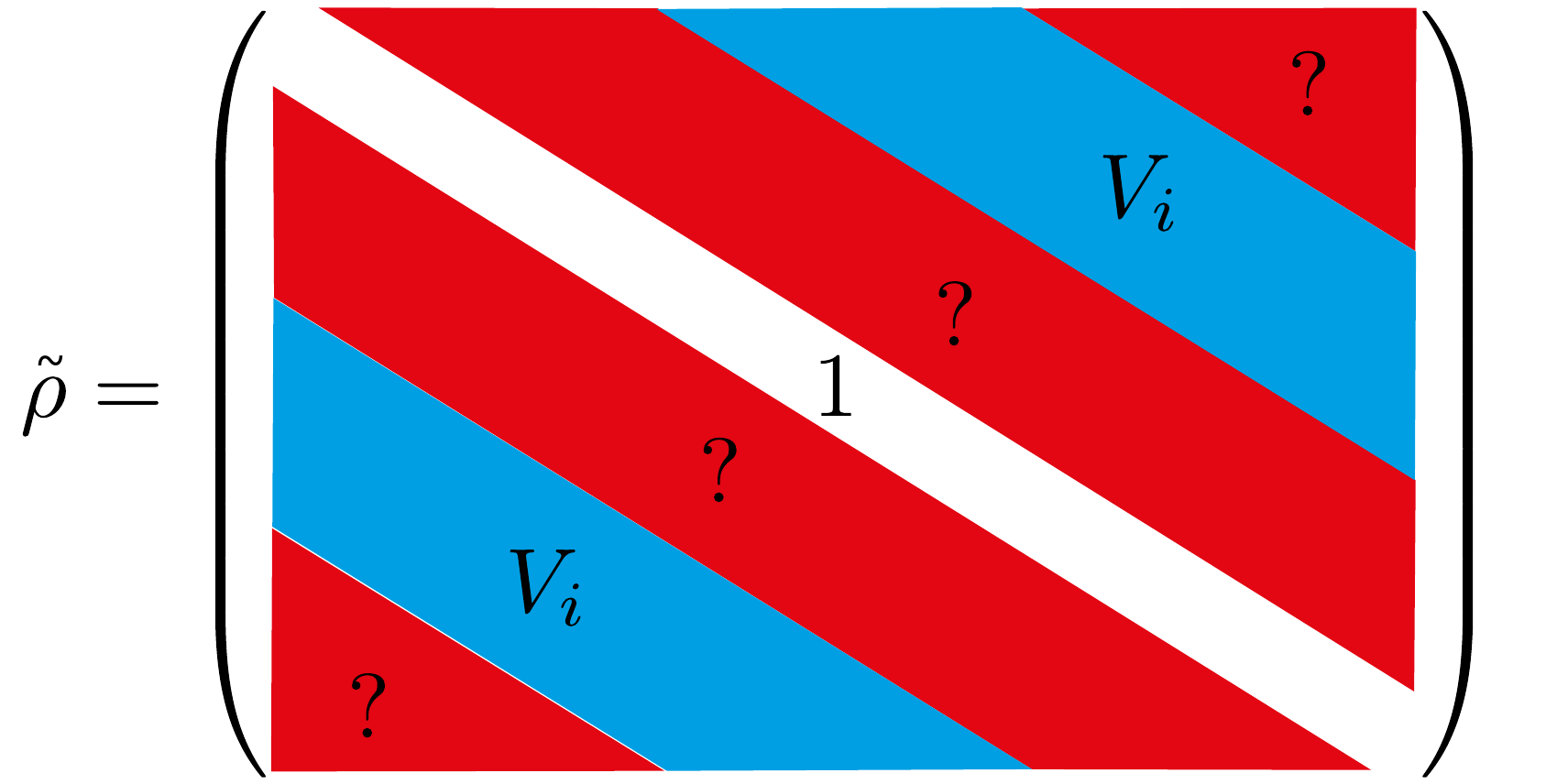}\\
\end{center}
\noindent Finally, to estimate the entanglement, we run the SDP procedure considering various values of the number of temporal modes up to $n=50$. As shown in \figurename{~\ref{fig_CW_vis}.(b)}, the data allows us to certify an entanglement of formation of $E_{oF} = 4.1(1)$ ebits, for $n=42$. This also certifies that our entangled state is of dimension at least $18 \times 18$. 

%Finally, note that the assumption on the equally distributed diagonal elements of $\tilde\rho$ can be relaxed, in which case the SDP procedure allows us to certify $E_{oF} = 4.1 (1)$ ebits, for $n=42$.

\emph{Conclusion.---} We have demonstrated the quantification and certification of high-dimensional photonic entanglement, based on sparse experimental data. First, using a time-bin encoding we certified an entanglement of formation of 2.09(7) ebits. Next moving to an implementation based on energy-time entanglement we could certify up to 4.1(1) ebits. This represents a considerable improvement over the largest values certified so far in any experiment; to the best of our knowledge, the highest value up until now was $E_{oF} = 1.2$ ebits in Ref.~\cite{Alexey2016}. This demonstrates that photonic systems encoded in temporal modes are ideally suited for the creation, certification, and quantification of high-dimensional entanglement. This opens promising perspectives for future applications in quantum information science.

\emph{Acknowledgments.} The authors would like to thank Hugo Zbinden for useful discussions. This work was supported by the Swiss national science foundation (SNSF 200021-149109), and the European Research Council (ERC-AG MEC). N.~B. acknowledges funding from the Swiss National Science Foundation (Starting grant DIAQ). M.~H. would like to acknowledge funding from the Swiss National Science Foundation (AMBIZIONE Z00P2-161351) and the Austrian Science Fund (FWF) through the START project Y879-N27.

\bibliography{bibliography_ebit}

\end{document}